\documentclass[a4paper,10pt,twoside]{cpc-hepnp}
\usepackage{multicol}
\usepackage{graphicx}
\usepackage{booktabs}
\usepackage{amssymb,bm,mathrsfs,bbm,amscd}
\usepackage[tbtags]{amsmath}
\usepackage{lastpage}
\usepackage{subfigure}

\begin{document}

\fancyhead[co]{\footnotesize Tian Li-Chao~ et al: Developments of a
2D Position Sensitive Neutron Detector}

\footnotetext[0]{Received 14 March 2009}

\title{Developments of a 2D Position Sensitive Neutron Detector}

\author{%
TIAN Li-Chao$^{1,2;1)}$\email{tianlc@ihep.ac.cn}%
\quad TANG Bin$^{2}$
\quad WANG Xiao-Hu$^{2}$
\quad LIU Rong-Guang$^{2}$ \\
\quad ZHANG Jian$^{2}$
\quad CHEN Yuan-Bo$^{2}$
\quad SUN Zhi-Jia$^{2;2)}$\email{sunzj@ihep.ac.cn}   \\
\quad Xu Hong$^{2}$
\quad YANG Gui-An$^{2}$
\quad ZHANG
Qiang$^{2,3}$ } \maketitle

\address{%
$^1$ Graduate University of the Chinese Academy of Sciences, Beijing, 100049, China\\
$^2$ Institute of High Energy Physics, Chinese Academy of Sciences, Beijing, 100049, China\\
$^3$ China University of Petroleum, Beijing, 102249, China \\
}

\begin{abstract}
Chinese Spallation Neutron Source (CSNS)£¬one project of the
12$^{th}$ five-year-plan scheme of China, is under construction in
Guangdong province. Three neutron spectrometers will be installed at
the first phase of the project, where two-dimensional position
sensitive thermal neutron detectors are required. Before the
construction of the neutron detector, a prototype of two-dimensional
$200mm\times200mm$ Multi-wire Proportional Chamber (MWPC) with the
flowing gas of Ar/CO$_{2}$ (90/10) has been constructed and tested
with the $^{55}$Fe X-Ray using part of the electronics in 2009,
which showed a good performance.

Following the test in 2009, the neutron detector has been
constructed with the complete electronics and filled with the
6atm.$^{3}$He + 2.5atm.C$_{3}$H$_{8}$ gas mixture in 2010. The
neutron detector has been primarily tested with an Am/Be source. In
this paper, some new developments of the neutron detector including
the design of the high pressure chamber, the optimization of the gas
purifying system and the gas filling process will be reported. The
results and discussion are also presented in this paper.

\end{abstract}



\begin{multicols}{2}

\section{Introduction}

Thermal neutron scattering techniques are playing an important role
in the diffraction experiments in determination of the molecular and
crystal structures in biology, condensed state physics and polymer
chemistry requiring a high flux neutron sources. Chinese Spallation
Neutron Source (CSNS), as the first spallation neutron source in the
developing countries, will be working at the beam power of 0.1 MW
and the neutron flux of $2.0\times 10^{16} cm^{-2}\cdot s^{-1}$.
Three spectrometers will be installed at the first phase of the
project. Efficient detectors with high position resolution, high
detection efficiency and low gamma sensitivity are required for
neutrons in the wavelength range from 1.8 $\textbf{\AA }$ to about 8
$\textbf{\AA }$. Because of the high cross-section for neutrons
absorption (~5330 b @ 1.8 $\textbf{\AA }$), the $^{3}$He gas is
widely used in many instruments\cite{lab1,lab2,lab3}. The detector
based on Multi-wire Proportional Chamber (MWPC) filled with $^{3}$He
gas can be built in large size, with relatively good energy and
position resolution, high efficiency and shows no radiation damage
compared to the solid state and scintillation detectors.

Usually£¬to meet the characteristics of high neutron detection
efficiency and high position resolution, the $^{3}$He based neutron
detectors are working in a high pressure state. The neutron detector
consists of a MWPC and a high gas pressure container filled with the
operating gas. To optimize the parameters of the MWPC, the main part
of the neutron detector, and guarantee that it can show excellent
performance in neutron test, it is necessary to construct a
prototype of the MWPC and test it by X-Ray with flowing gas firstly.
Good results have been achieved in the X-Ray test in 2009, then a
high pressure container was constructed and the MWPC was fixed in
it. Later on, the operating gas was filled with good gas tightness
and the neutron detector equipped with the complete electronics was
tested with the Am/Be source at the Institute of High Energy Physics
(IHEP).

\section{The MWPC prototype \cite{lab4}}

A prototype of a two-dimensional MWPC with atmospheric pressure and
flowing gas for X-Ray has already been constructed prior to the
neutron detector. The geometry of the two-dimensional MWPC is shown
in Fig.~\ref{fig:MWPC}. The MWPC with $200mm\times200mm$ sensitive
area is of conventional design with a cathode plane, an anode plane
and two orthogonal readout planes symmetrically located about the
central anode plane. The cathode is made of a thin metal foil pasted
on the window. The anode plane is made up of 15 $\mu$m diameter
gold-plated tungsten wires (with the tension of 25 g per wire) with
an inter-wire spacing of 2 mm. The X readout plane is made up of 50
$\mu$m diameter gold-plated tungsten wires (with the tension of 40 g
per wire) with an inter-wire spacing of 1 mm. Every four wires are
connected together to form one readout-strip. The direction of the
readout wires is parallel to the anode wires. The Y readout grid is
made up of 1.6 mm wide copper strips in the orientation orthogonal
to the anode wires with an inter-strip spacing of 2 mm. Every two
copper strips are connected together to form one readout strip.

\begin{center}
\includegraphics[width=6cm]{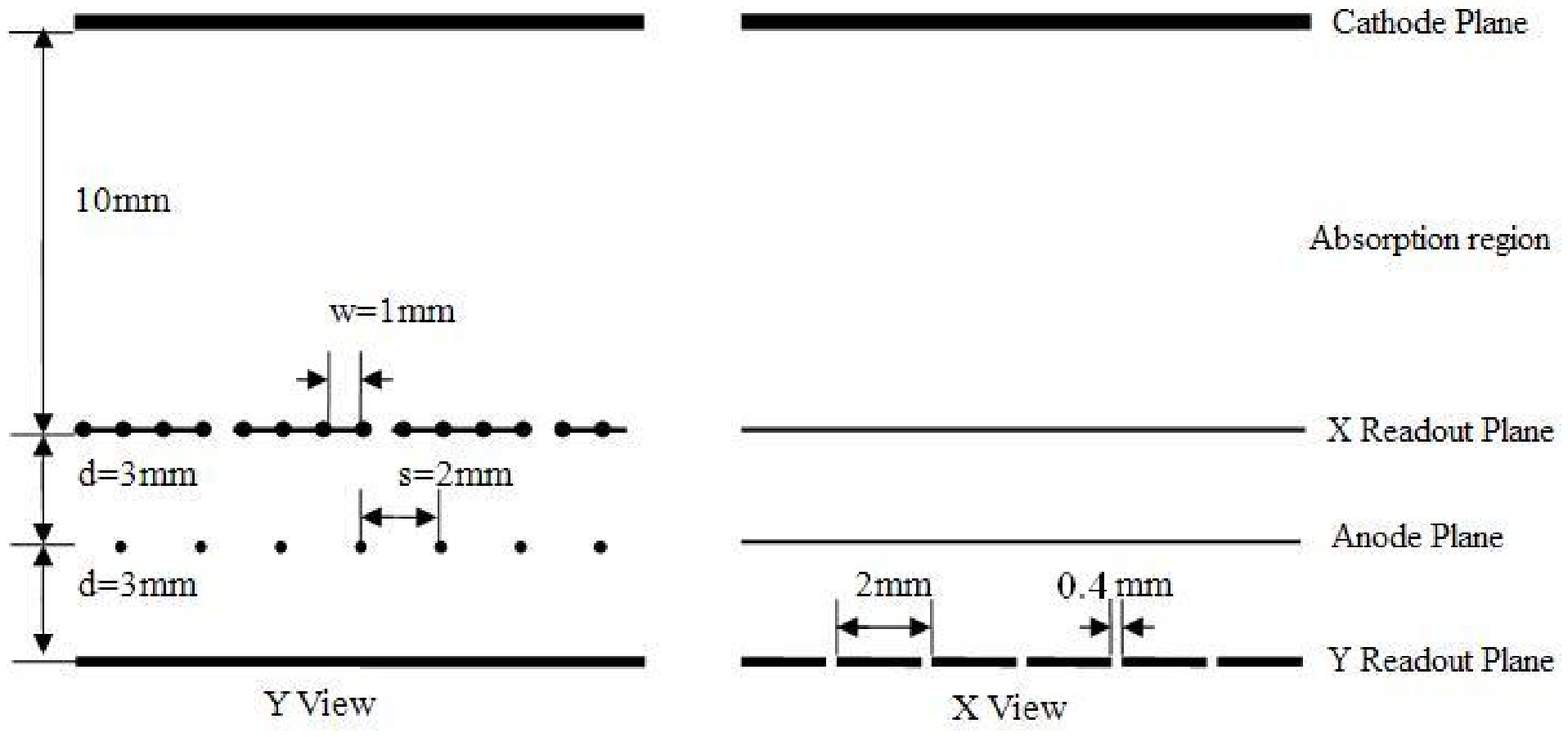}
\figcaption{\label{fig:MWPC} layout of the planes of the MWPC}
\end{center}

Tested by $^{55}$Fe 5.9keV X-ray in 1 atm. Ar/CO$_{2}$ (90/10)
mixture with part of the electronics, an energy resolution (FWHM) of
about 23\% for $^{55}$Fe 5.9keV X-ray, and a spatial resolution
(FWHM) of about 210$\mu$m along the anode wire direction were
obtained (See Fig.~\ref{fig:performance}). The successful
construction of the prototype laid the foundation for the neutron
detector construction.

\begin{center}
\makeatletter\def\@captype{figure}\makeatother \subfigure[imageing
ability]{
    \label{fig:performance:a}
    \includegraphics[height=3cm]{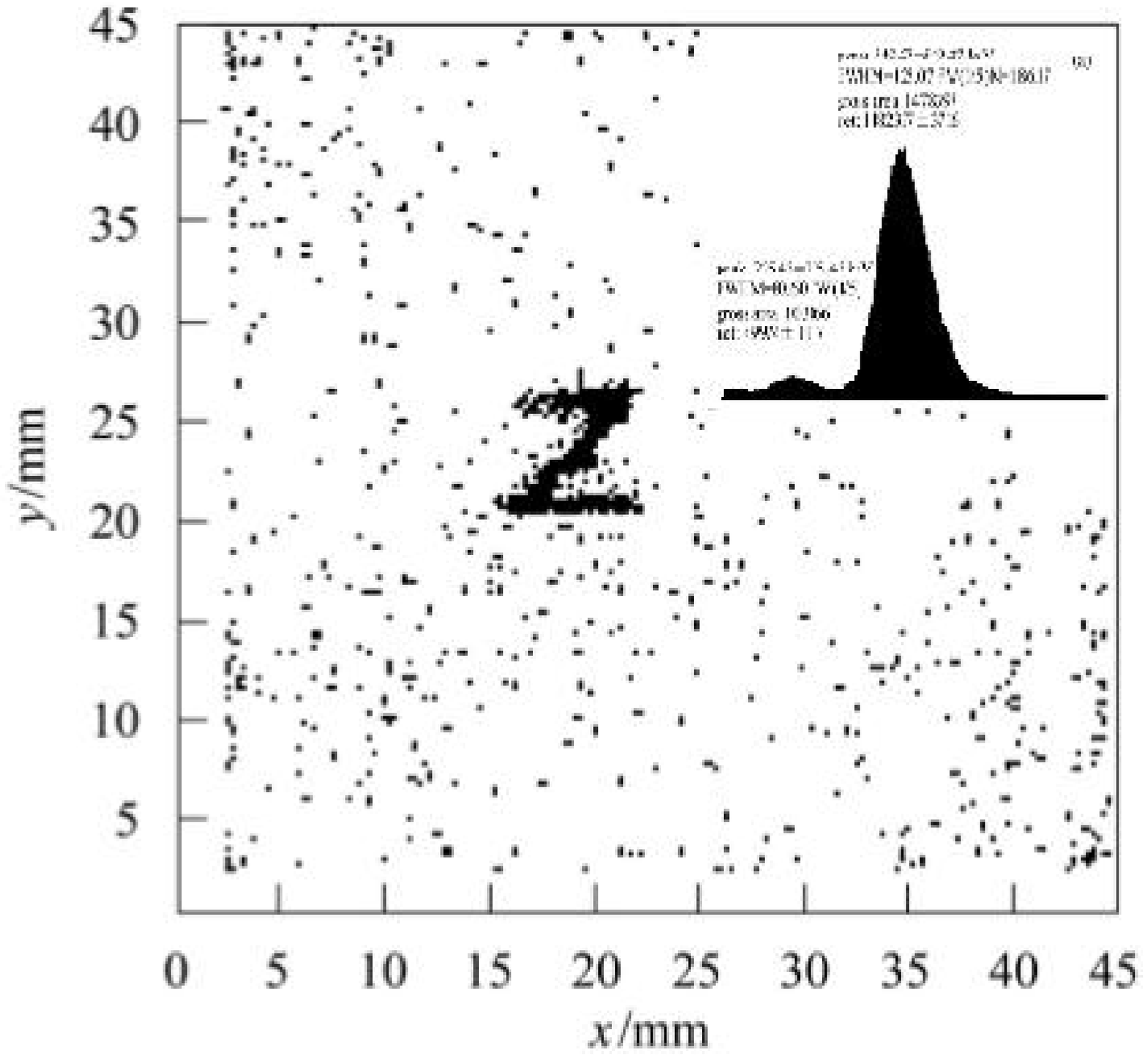}}
\makeatletter\def\@captype{figure}\makeatother
\subfigure[position resolution]{
    \label{fig:performance:b}
    \includegraphics[height=3cm]{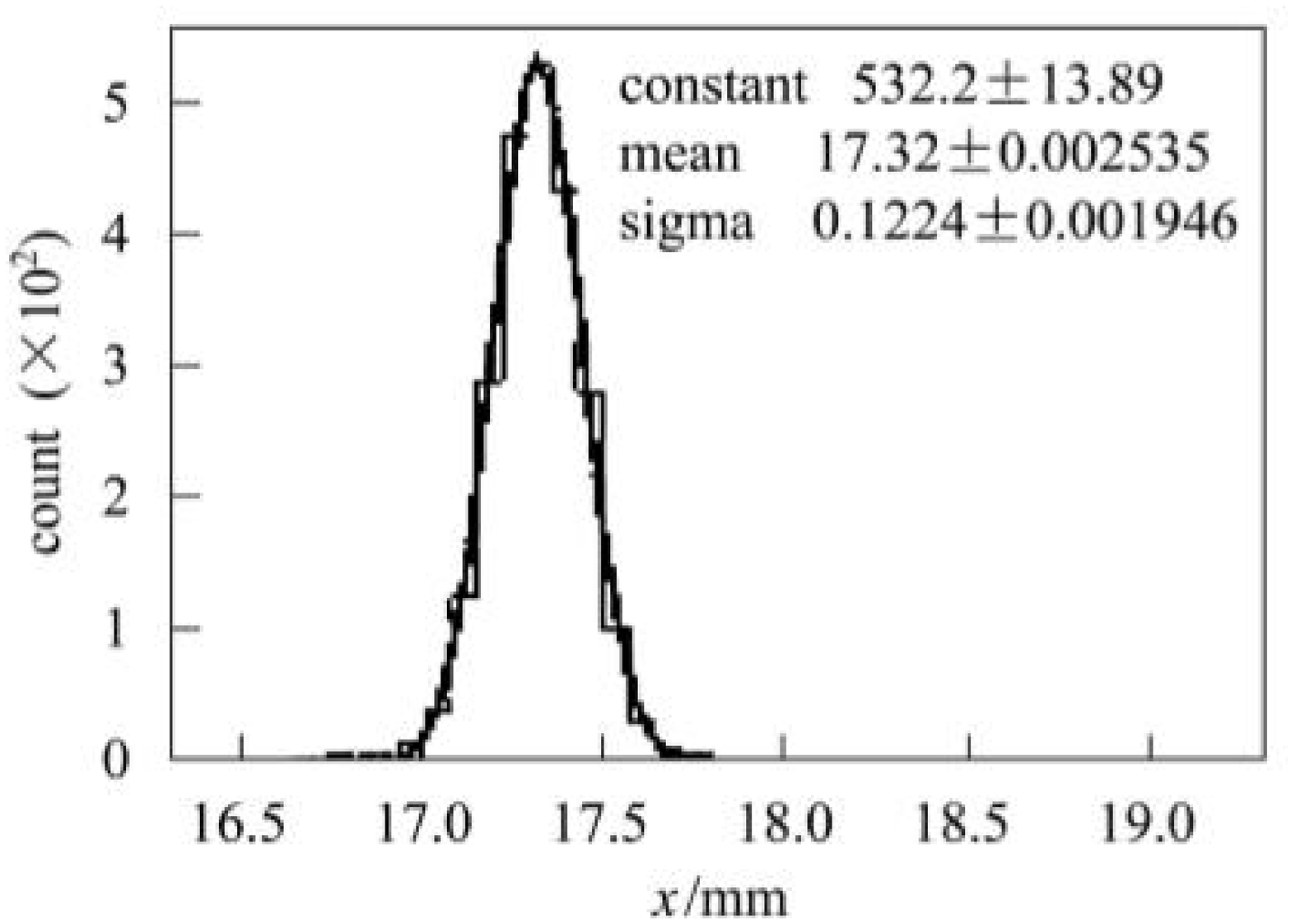}}
\caption{performance of the prototype} \label{fig:performance}
\end{center}

\section{Developments}
The main structure of the thermal neutron detector is the same as
the prototype (See Fig.~\ref{fig:MWPC}). The neutron absorption by
$^{3}$He induces a fission reaction and emission of two charged
particles, one triton and one proton, in opposite direction with a
total kinetic energy of 764 keV which induces the primary ionization
in the gas. The position of the neutron is considered to be the
centroid of the primary electrons associated with the nuclear
reaction point. But actually the ionization centroid is displaced to
the point of 0.4 times of the proton moving range apart from the
nuclear reaction point \cite{lab5}. To meet the characteristics of
high neutron detection efficiency and high position resolution, the
wire planes should be sealed in a high pressure chamber with 6atm.
$^{3}$He as detection medium and 2.5atm. C$_{3}$H$_{8}$ as stopping
gas. A high pressure chamber has been designed, the gas purifying
system was updated and all the electronics were tested before the
performance test with the Am/Be source.

\subsection{High pressure chamber}
As will be working under high pressure, all the wire planes should
be fixed in a high pressure container, which consisted of a circle
front cover made of aluminum alloy 6061-T6 and a circle back-plate
made by stainless steel (See Fig.~\ref{fig:photo}). Double "O" shape
rings were used to seal the container. A 9 mm thick window with area
of $210mm\times210mm$ was constructed in the center of the front
cover to minimize the neutron scattering . The 25 mm thick
back-plate provided structural rigidity under the maximum gas
pressure of 10 atm. The anode signals and the cathode signals were
transferred to the readout electronics via flanges deployed on the
back-plate. A gas purifying system, consisting of a pump and a
filter, was also included, so as to guarantee the purity of the
operating gas \cite{lab6}.

\begin{center}
\includegraphics[width=5cm]{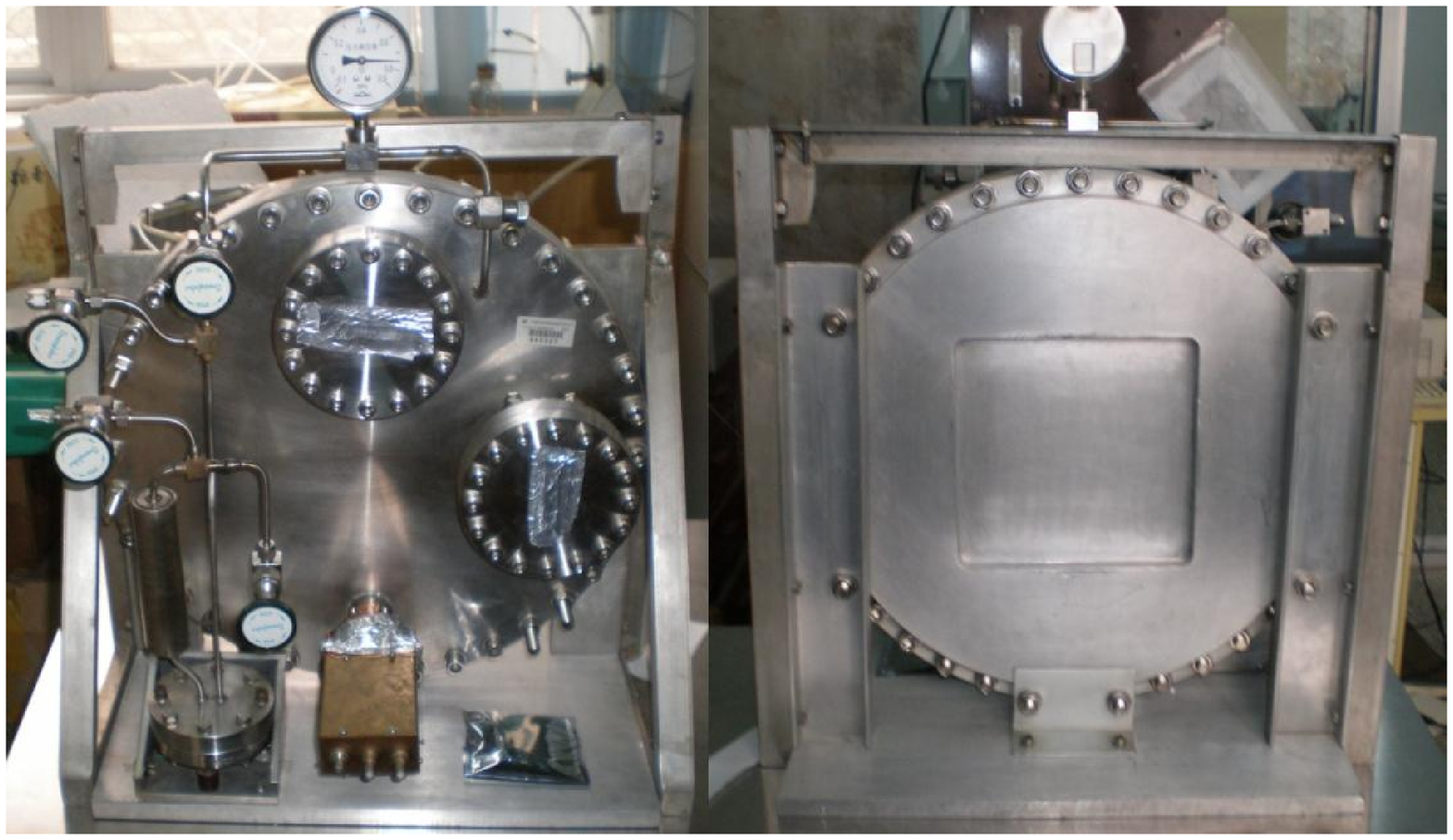}
\figcaption{\label{fig:photo} the pictures of the high gas pressure
chamber}
\end{center}

\subsection{Gas filling}
The gas purifying system has been updated since there was a little
gas leakage and the gas filling process has been completed in
Beijing Nuclear Instrument Factory in February, 2010. Some
unnecessary tubes were removed to reduce the gas leaking
probability. To test the gas tightness, 9 atm. of $^{4}$He gas has
been filled in the chamber and kept for about two months without
obvious leakage. Before being filled with the operating gas, the
chamber was first heated to 60$^{\circ}$C and evacuated until a
stable vacuum was obtained (about $3\times 10^{-4}$ Pa) to release
the residual gas in the materials. Kept the vacuum state for 48
hours, and then flowed the pure argon through the chamber for 5
hours, so as to take away the leftover gas in the chamber. Secondly,
the chamber was evacuated to vacuum again. Finally,
$^{3}$He/C$_{3}$H$_{8}$ mixture was filled.
Fig.~\ref{fig:pressure:a} shows the changes of the gas pressure and
the temperature after the 8.5 atm. of operating gas filled in 45
days. The fluctuation was caused by the temperature.
Fig.~\ref{fig:pressure:b} shows the ratio of the calculated
pressure, corrected by temperature, to the initial pressure in each
day, which are within the accepted reading error limits.

\begin{center}
\makeatletter\def\@captype{figure}\makeatother \subfigure[Pressure
and temperature varieties along time]{
    \label{fig:pressure:a}
    \includegraphics[width=3.8cm]{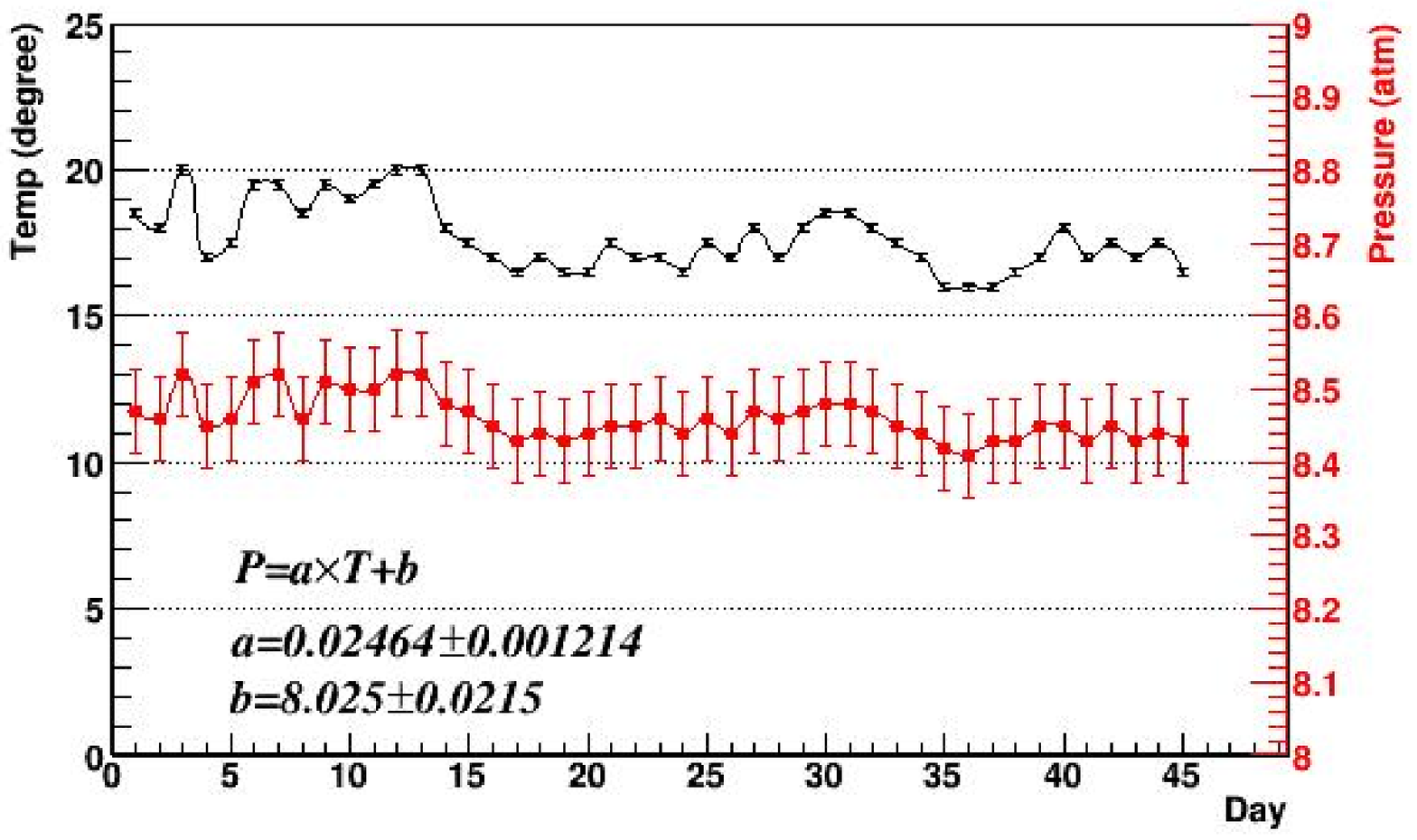}}
\makeatletter\def\@captype{figure}\makeatother \subfigure[Ratio of
calculated pressure to the initial pressure]{
    \label{fig:pressure:b}
    \includegraphics[width=3.8cm]{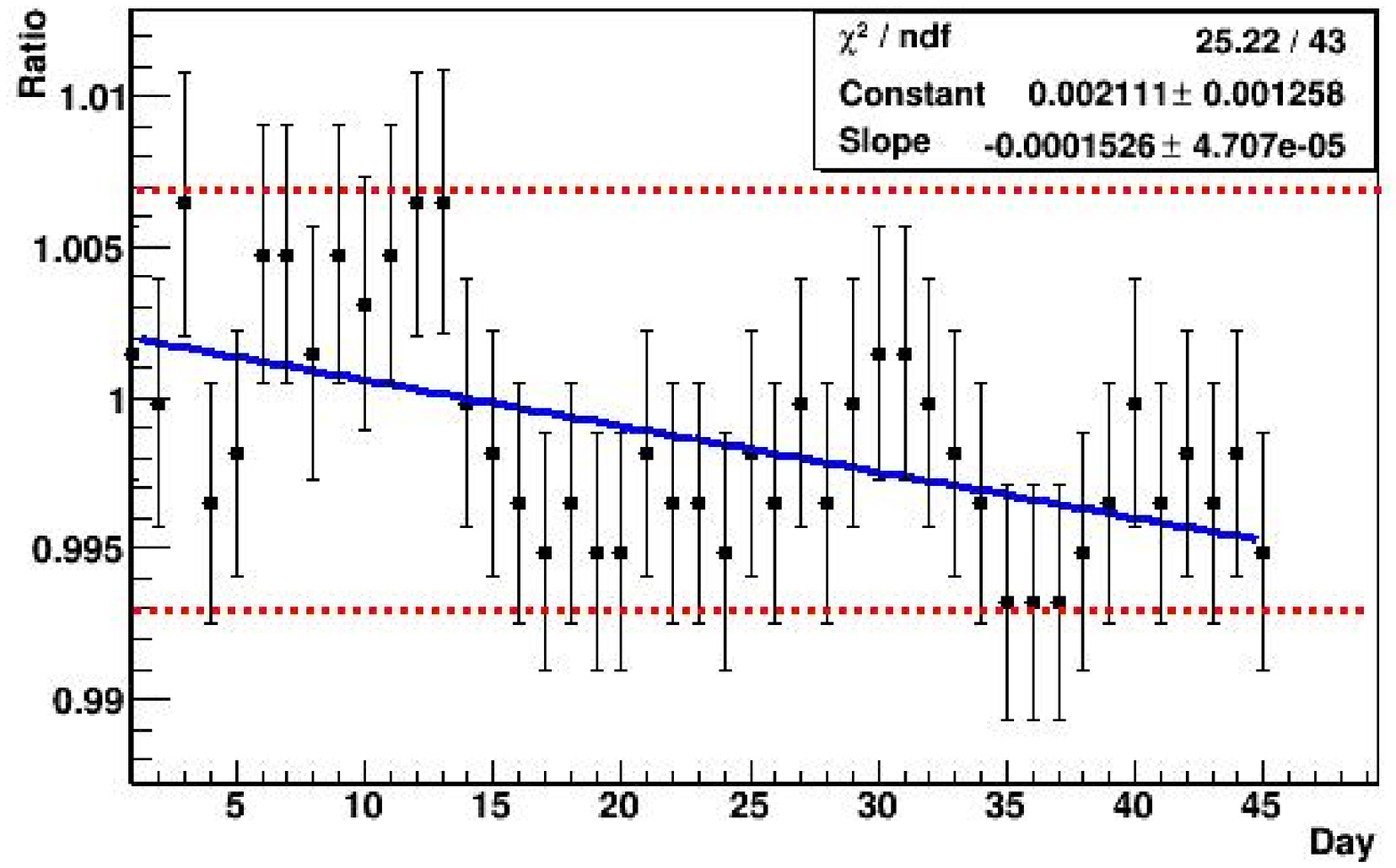}}
\caption{Pressure variety of the Chamber} \label{fig:pressure}
\end{center}

Presumed it is an exponential leaking, the ratio changes with time
as the function~(\ref{eq:leakege}), so the leakage lifetime is
$T_{0}=18$ years. It means that 18 years later the gas pressure will
be 1/e of the initial pressure. Since the C$_{3}$H$_{8}$ molecules
are much larger than $^{3}$He atoms, the main leaking gas will be
$^{3}$He. There will be effects on the detection efficiency and the
gas gain as the gas leaking, which are shown in Table~\ref{tab1},
assuming the detector is working with the same high voltage. So, in
actual running to keep the uniformity of the gas gain the operation
high voltage should be decreased as time.

\begin{eqnarray}\label{eq:leakege}
R=e^{-0.0001526t+0.002111}
\end{eqnarray}

\begin{center}
\tabcaption{ \label{tab1}  The effects on the detector performance
caused by the gas leakage (simulated by Garfield with same high
voltage).} \footnotesize
\begin{tabular*}{80mm}{c@{\extracolsep{\fill}}ccc}
\toprule Elapsed time   & Detection efficiency  & Relative change of Gain \\
\hline
Initial state\hphantom{0}        & \hphantom{0}75\%           & 1 \\
6 years ($\textit{T}$$_{0}$/3)\hphantom{0}  & \hphantom{0}59\%           & 196\% \\
9 years ($\textit{T}$$_{0}$/2)\hphantom{0}  & \hphantom{0}50\%           & 241\% \\
18 years ($\textit{T}$$_{0}$) \hphantom{0}  & \hphantom{0}25\%           & 407\%\\
\bottomrule
\end{tabular*}
\end{center}

\subsection{Electronics}
The location of the neutron is assumed to be the gravity center of
the induced charges read out from the cathode strips. There are two
aspects which should be taken into account during the electronic
design. First of all, the detector should own a pretty good
performance in n/$\gamma$ discrimination. The $\gamma$ rays can be
discriminated from neutrons due to the differences of energy deposit
in the gas. Secondly, we have to obtain the distribution of the
induced charge on the cathode strips to get the center of gravity to
reconstruct the position of the neutrons. A total of 100 individual
position readout channels, 50 in each direction, and one anode
readout channel are developed by the electronic group in IHEP. The
total charge collected on the anode wire is used to get the energy
loss information to trigger the readout. The analogue signal from
the charge-sensitive preamplifier is directly converted to digital
signals by a 10 bit 40 MHz FADC. Then the peak finding circuits
based on FPGA, which can handle 16 channels at the same time, obtain
the charge of each readout channel and deduct the baseline, and then
keep the results in a local buffer (See Fig.~\ref{fig:eletronics}).
All the electronics after the preamplifiers are built on the VME
daughter card. The DAQ can read the buffer through the VME bus and
save the raw data in a local hard disk for the offline analysis. All
the electronics has been calibrated together with the VME crate,
which showed a good long-term stability.

\begin{center}
\begin{minipage}[t]{0.45\linewidth}
    \includegraphics[width=3cm]{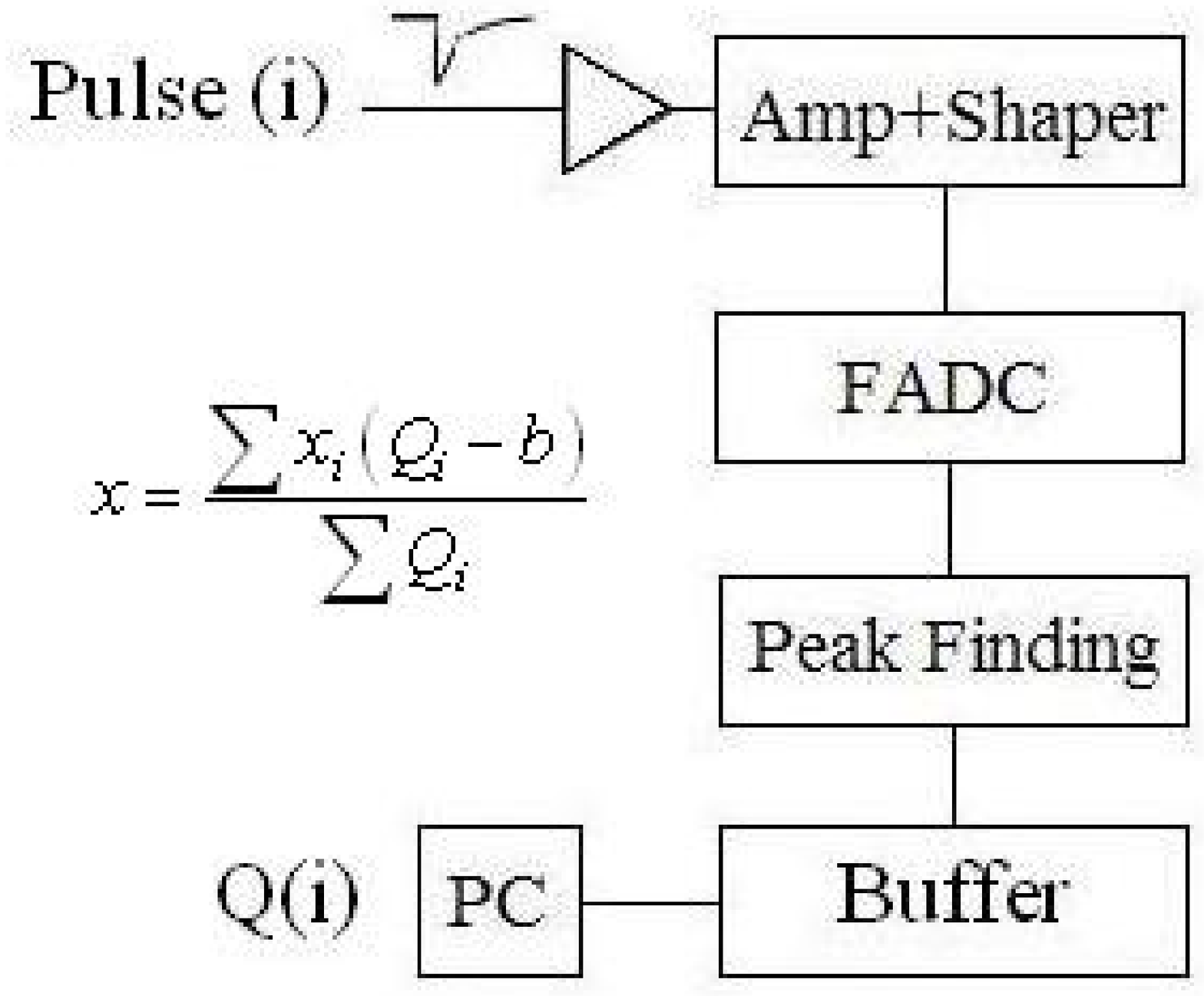}
    \figcaption{\label{fig:eletronics} Block diagram of the electronic}
\end{minipage}
\hspace{3ex}
\begin{minipage}[t]{0.45\linewidth}
    \includegraphics[width=3cm]{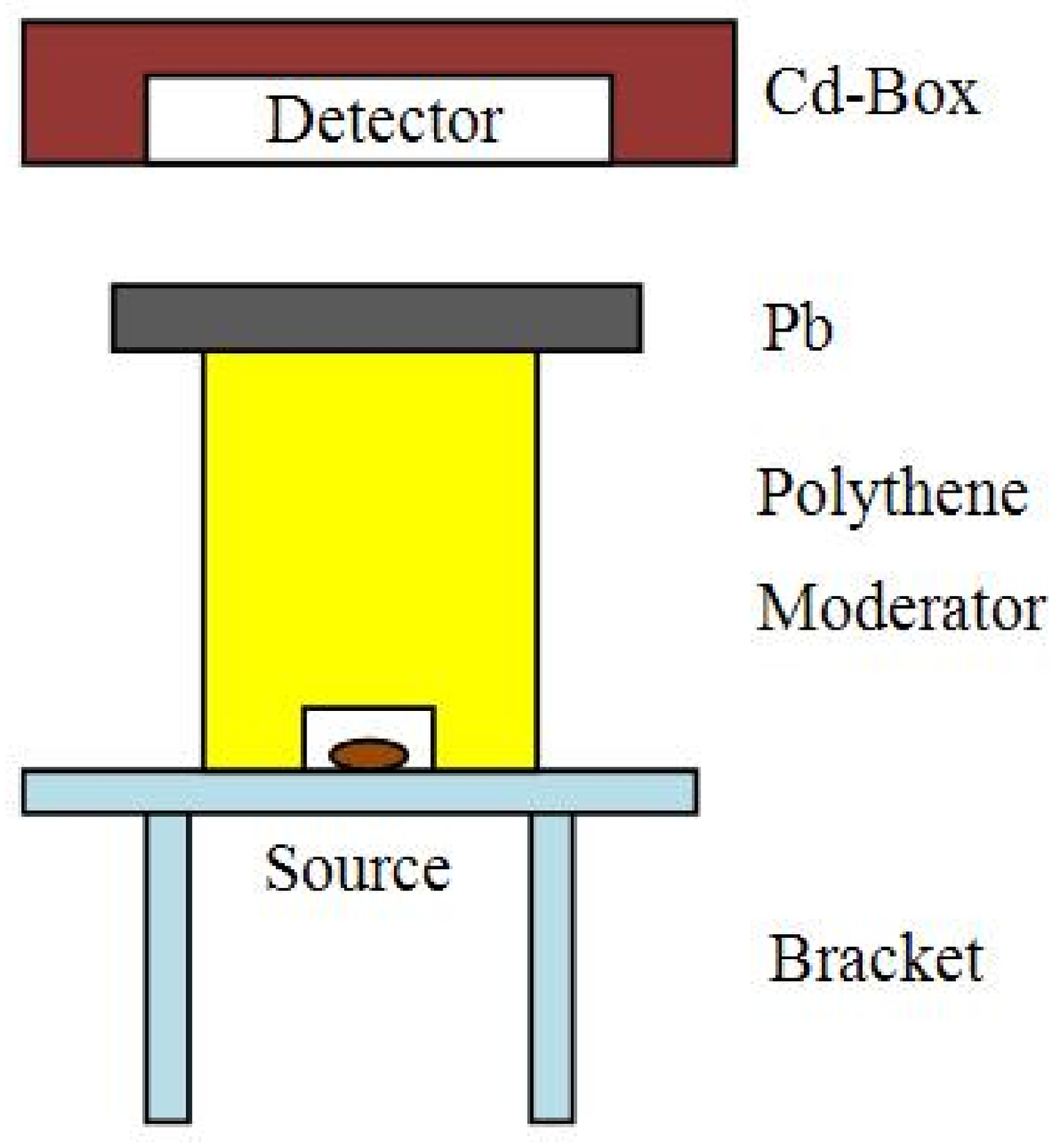}
    \figcaption{\label{fig:exp_setup} schematic map of the experimental equipment}
\end{minipage}
\end{center}

\section{Test with the Am/Be source}
\subsection{Experimental Setup}
The neutron detector was firstly tested with an Am/Be source, which
was sufficient for uniform irradiation testing, but not powerful
enough to generate collimated beams for high spatial resolution
testing. The experimental equipment was set up as
Fig.~\ref{fig:exp_setup}. In our experiment, the fast neutrons
emitting from the Am/Be source were moderated through a 40 cm
thickness polythene block firstly and then went through a 12 mm
thick Pb (as the shield of the gamma rays). Finally, the moderated
neutrons entered the detector with a cadmium mask placed in front of
the detector window. The whole detector was put in a Cd-box to
absorb the surrounding thermal neutrons.

\subsection{Results}

\noindent A) Spectrum

The energy spectrum is the main basis in the n/$\gamma$
discrimination. The energies deposited in the detector under
different testing conditions are shown in Fig.~\ref{fig:spectrum}.
The spectrum (a) was tested with the $^{137}$Cs gamma ray source;
the spectrum (b) was tested with an Am/Be source with a moderator
and Pb in front of the window shown in the Fig.~\ref{fig:exp_setup};
the spectrum (c) was tested under the same condition with (b) but
with 4 mm thick cadmium more in front of the window to absorb the
thermal neutrons from the source. Three spectrums had been
normalized by the run time. The peaks on the right side corresponded
to the energy disposed by neutrons, while the middle ones near the
200 channel corresponded to the gamma rays. The left peak was caused
by the fake triggers due to the electronic noise.

With appropriate thresholds selected, the gamma ray could be
discriminated from the neutron easily through the comparing of the
spectrum (a) with (b). But there were many upper-Cd neutrons and
high energy gamma rays (together about 17\% of the thermal neutrons)
which could still penetrate the cadmium and be captured in the
detector from the comparing of the spectrum (b) with (c). They were
the main background in the thermal neutron test.

\begin{center}
\includegraphics[width=5cm]{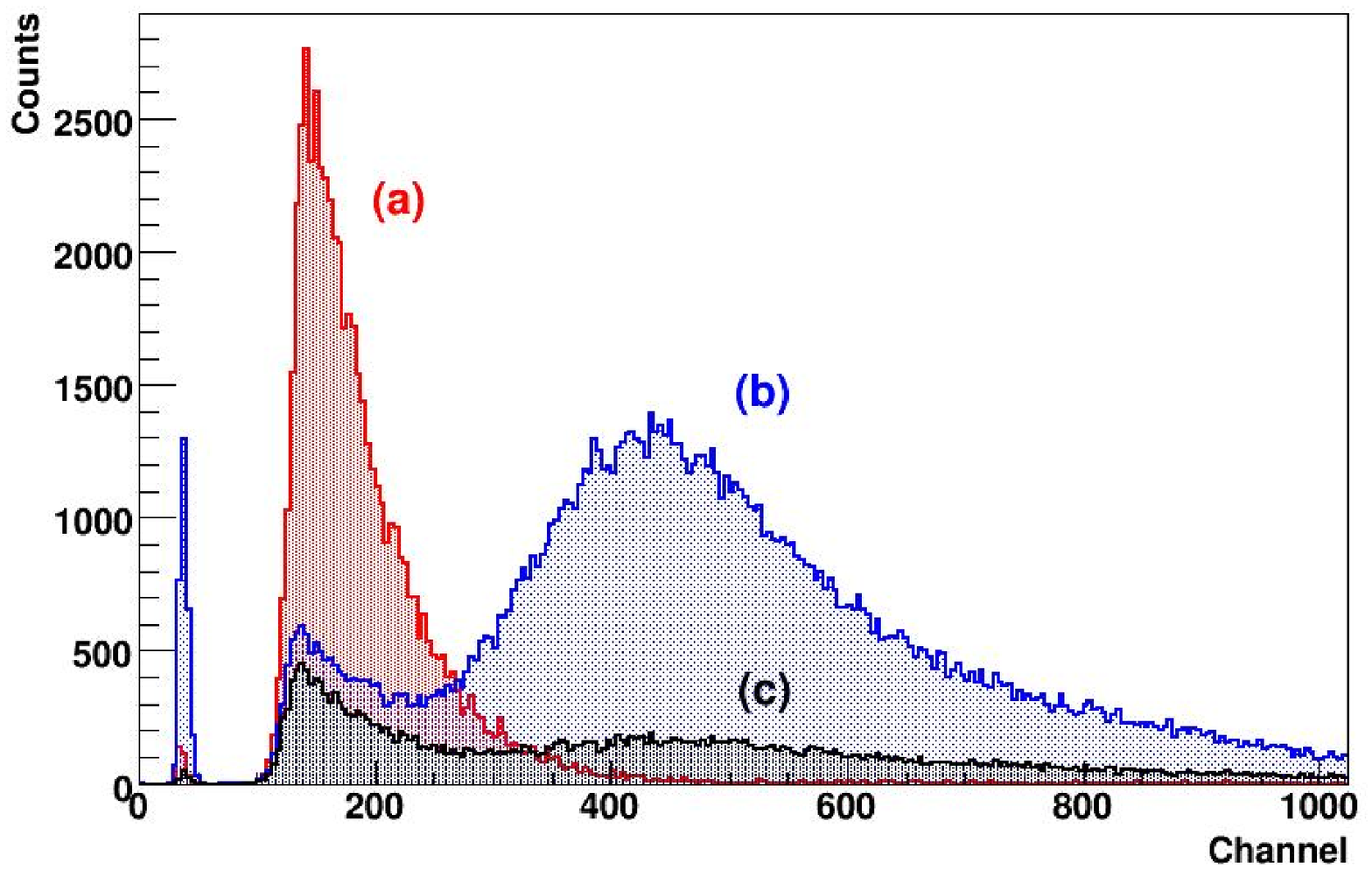}
\figcaption{\label{fig:spectrum} Energy spectrums under different
test conditions }
\end{center}

\noindent B) 2D Imaging

A two-dimensional image was obtained by placing a mask (Shown in
Fig.~\ref{fig:h_image}(a)) in front of the detector's window. The
mask with 2mm thick was made up of a cadmium plane with a hole of
the character "H". In order to get a sharp image we had to make the
strokes of the character a little wider, due to the low intensity of
the source and the high neutron scattering in the moderator. To
collect enough neutron events the whole system had to be taking data
for at least 10 hours. The charge collected on the anode, the number
of fired strips, the maximum charge and the sum of the charge
induced on fired strips were used to discriminate the gamma rays and
neutrons in the analysis. This image (Shown in
Fig.~\ref{fig:h_image}(b)) could also illustrate the good uniformity
of response.

\begin{center}
\includegraphics[width=8cm]{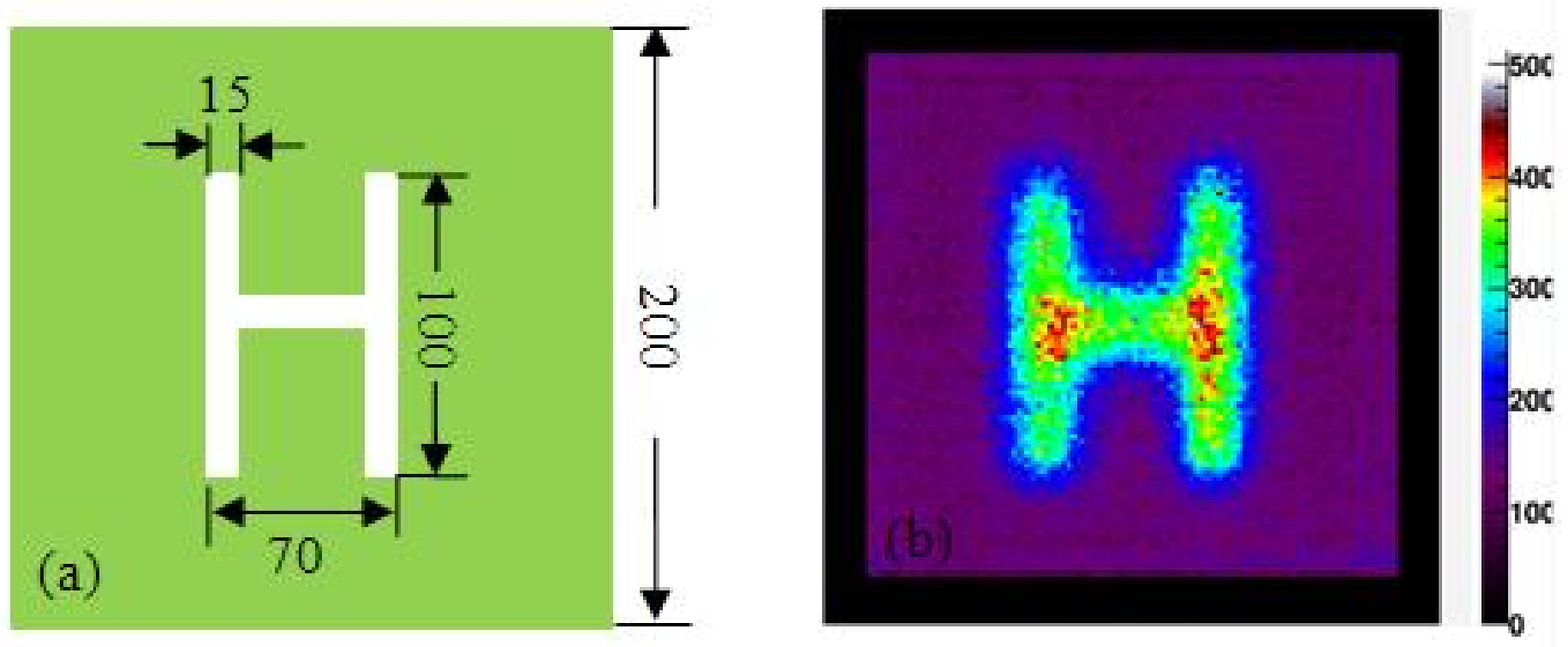}
\figcaption{\label{fig:h_image} Cd mask (a) and the 2D image (b) }
\end{center}

\noindent C) The position linearity

The position resolution was tested in the direction paralleling to
the anode wires (Y direction) at variable positions. Due to the low
flux of Am/Be source, only a rough position resolution about 4.9 mm
(sigma) was achieved. The position linearity was tested and the fit
function of the measured positions to the actual positions was
$\textit{y=1.0117x+0.0811}$. (See Fig.~\ref{fig:linearity}). The
linear correlation coefficient was 0.9998, which showed a good
linearity.

\begin{center}
\includegraphics[width=5cm]{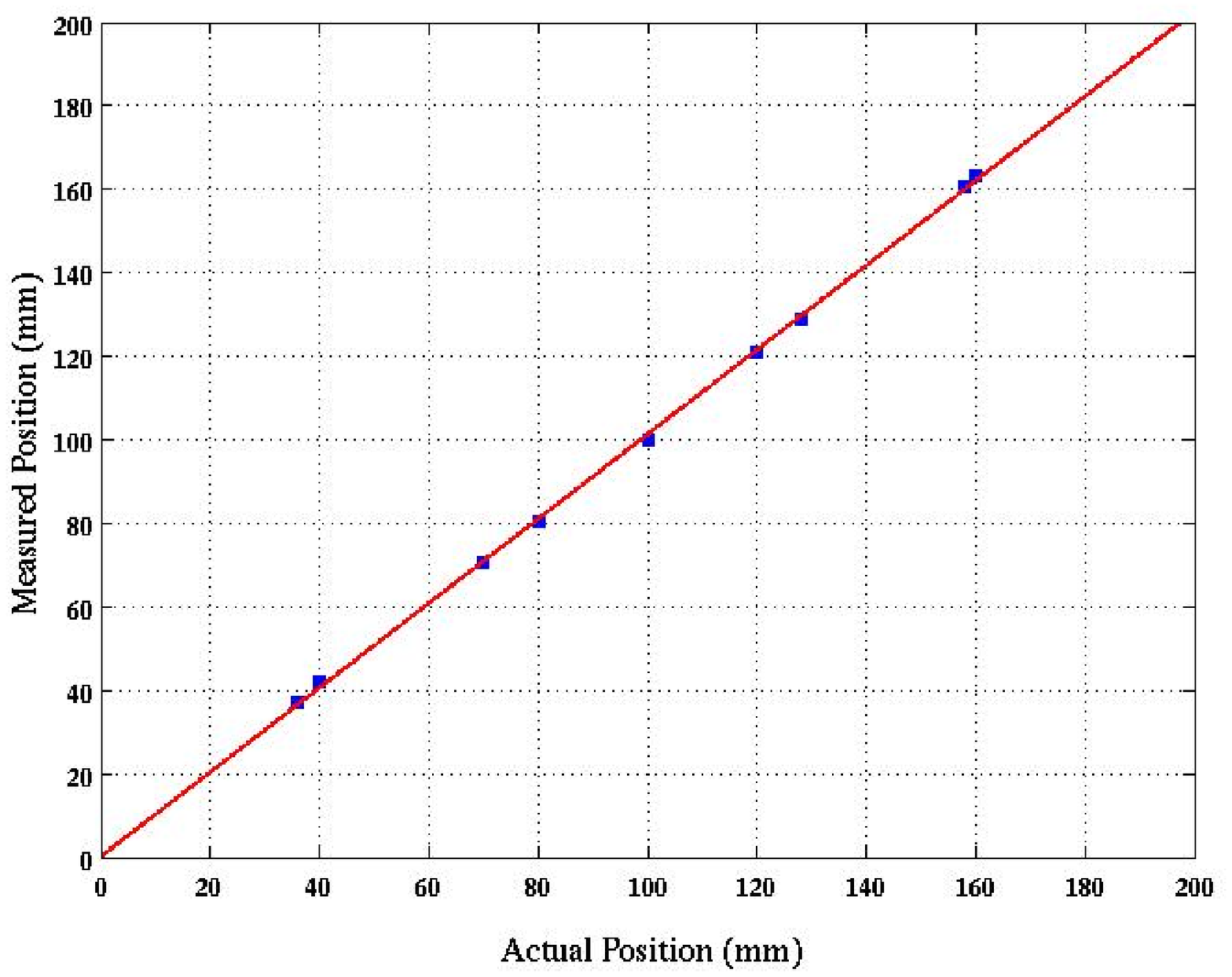}
\figcaption{\label{fig:linearity} the position linearity of the
detector}
\end{center}

\subsection{Discussion}
The spatial resolution of the neutron detector is mainly determined
by the proton moving range in the operation gas and the intrinsic
position resolution of the MWPC. The proton moving range is
determined primarily by the propane pressure - it is about 1.43 mm
for our choice of 2.5 atm. from the SRIM simulation. The approximate
intrinsic position resolution of MWPC was 210 $\mu$m from the
$^{55}$Fe test. So, the theoretical neutron position resolution
should be the quadratic addition of the error caused by the proton
moving range and the intrinsic position resolution, which is better
than 1.45 mm (FWHM) according to the equation~(\ref{eq:resolution}).

\begin{eqnarray}\label{eq:resolution}
\sigma _{n}= \sqrt{\sigma _{g}^{2}+\sigma _{i}^{2}}
\end{eqnarray}

Where, $\sigma _{n}$ is theoretical neutron position resolution;
$\sigma _{i}$ is the intrinsic resolution from the X-Ray test;
$\sigma _{g}$ is the error caused by the proton moving range.

The reasons why the position resolution in our test was not so good
as the expected result were as follows:

\noindent i. The high background.

The background, mainly the upper-Cd neutrons and high energy gamma
rays from the surrounding, was too high even though the detector was
put in a Cd-Box since there were other neutron sources and gamma
sources in the lab. The signal to noise ratio was very low. The
intensity of thermal neutron from the moderator was not high enough
that we had to take data for long time.

\noindent ii. Hard to collimate the neutrons.

The neutrons through the moderator could be considered as a plane
source in front of the detector window since the neutron emitting
direction, the angle between the emitting direction and the normal
line of the detector's window, was about from $0\sim 90$ degree
shown in Fig.~\ref{fig:angle}. So, most of the thermal neutrons
could not go into the detector perpendicular to the window, which
made the edge of the character "H" not clear enough.

\begin{center}
\includegraphics[width=5cm]{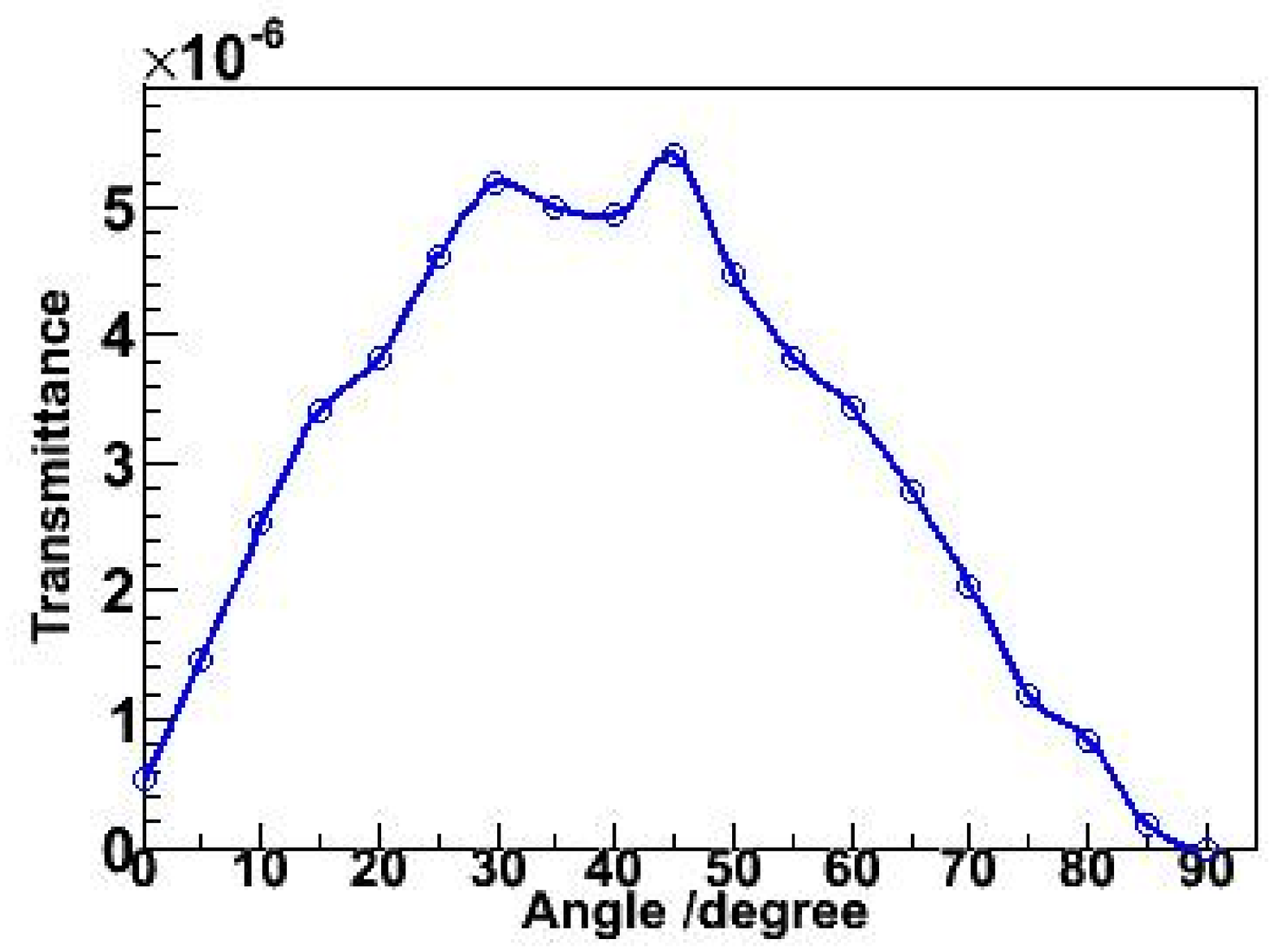}
\figcaption{\label{fig:angle} the thermal neutron emitting angle
distribution form the moderator (simulation result)}
\end{center}

\section{Conclusion}
The two-dimensional thermal neutron detector has been constructed
with the complete electronics system and filled with the 6 atm.
$^{3}$He+2.5 atm. C$_{3}$H$_{8}$ gas mixture in 2010. The leakage
lifetime of the chamber was about 18 years after the updates of the
gas purity system. The whole electronic system had good long-term
stability. The detector had a good ability in n/$\gamma$
discrimination. As for the low intensity Am/Be source being not
powerful enough to generate collimated beams, only a 2D image was
obtained with the position resolution of 4.9 mm in sigma. But good
position linearity was achieved also. A much better spatial
resolution is expected to be achieved on the collimated thermal
neutron source where there is much higher flux thermal neutrons.

\end{multicols}

\vspace{-1mm}
\centerline{\rule{80mm}{0.1pt}}
\vspace{2mm}

\begin{multicols}{2}

\end{multicols}

\clearpage

\end{document}